\documentclass[twocolumn]{aastex631}
\usepackage{graphicx}
\usepackage{natbib}
\usepackage{placeins}

\begin{document}

   \title{Disc Candidates in IC 2395: A WISE Survey of the Kinematically Confirmed Membership}

   \author[0009-0003-8540-0264]{Dan Rose}
   \affiliation{Independent Researcher, Ocean Grove, Victoria, Australia}

\begin{abstract}
Circumstellar disc dissipation timescales constrain the window available for planet formation. At $\sim$9\,Myr, \object{IC\,2395} lies at a critical epoch in primordial disc evolution, yet no kinematically selected disc census exists beyond the inner cluster core.

A wide-field survey of \object{IC\,2395}, extending to a 2\degr\ radius, establishes a kinematically selected disc census and evaluates the frequency of planet-forming environments in the outer cluster field.

Using a high-purity catalogue of 173 members identified via Gaia DR3 kinematics, mid-infrared excesses are identified using AllWISE photometry. Candidates are classified into two tiers: Cross-validated (corroborated by Gaia variability) and Single-band detections.

Twenty-one disc candidates are identified, 90\% of which lie beyond the footprint of prior Spitzer surveys. The candidates are concentrated among low-mass members (0.31--0.68\,M$_\odot$), with no significant correlation between mass and excess (Spearman $\rho = +0.07$, $p = 0.84$). A secure disc fraction of $4.0 \pm 1.5$\% is established by the Cross-validated subsample, independently confirmed by Gaia DR3 YSO variability classification; a photometric $3\sigma$ significance cut yields a consistent estimate of $2.9 \pm 1.3$\%. An upper bound of $12.1 \pm 2.5$\% is derived from all 21 W1$-$W2 excess candidates, with a background-subtracted value of $\sim$10.1\%. Gaia DR3 epoch photometry reveals diverse variability morphologies, including a dramatic dipper and multiple bursters, confirming ongoing magnetospheric accretion.

By extending the survey radius, this work demonstrates that infrared excess sources indicative of inner-disc emission in \object{IC\,2395} are more abundant and widely distributed than previously recognised. The survival of these discs in the lower-density outer field provides a well-characterised sample for studying the final stages of disc evolution and planet formation.
\end{abstract}


\section{Introduction}
\label{sec:intro}
The circumstellar discs that surround young pre-main-sequence stars are the
birthplaces of planets. The fraction of stars retaining primordial discs as a
function of age --- the disc fraction --- places direct observational constraints
on the timescales available for planet formation and the processes driving disc
dissipation. Surveys of young stellar clusters have established that primordial
disc fractions decline approximately exponentially from $\sim$80\% at 1~Myr to
fewer than 10\% by 8--10~Myr, with a characteristic dissipation timescale of
2--3~Myr \citep{mamajek2009, fedele2010, richert2018}. This rapid decline reflects a
combination of accretion onto the central star, photoevaporation by stellar
radiation, and dynamical interactions within the disc. The timescale is also
strongly mass-dependent: discs around lower-mass K- and M-type stars survive
systematically longer than those around higher-mass solar-type and earlier stars
\citep{luhman2012, ribas2015}.

At $\sim$9\,Myr, the young open cluster IC~2395 sits at a critical epoch in
primordial disc evolution --- old enough that most discs have dissipated, yet
young enough that surviving discs around lower-mass members remain physically
plausible and scientifically significant. IC~2395 is located in the
\object{Vela}/\object{Puppis} star-forming complex at a distance of $\sim$750~pc, with an
estimated age of $\sim$9\,Myr and modest extinction ($A_V = 0.6$~mag;
\citealt{Rose2026}). The only prior infrared disc census of IC~2395 was
performed by \citet{Balog2016} using Spitzer IRAC and MIPS photometry,
identifying 18 Class~II (6.5\%) and 8 transitional disc (2.9\%) candidates
within the central 44\arcmin\ $\times$ 44\arcmin\ Spitzer survey footprint. That survey utilised statistical membership models which, prior to Gaia DR3, were subject to field star contamination. The relatively small footprint also leaves the outer cluster population
entirely uncharacterised.

The Gaia mission has transformed the membership analysis of open clusters.
Precise proper motions and parallaxes from Gaia~DR3 \citep{GaiaCollaboration2023} enable
kinematically clean membership catalogues extending well beyond the angular
radii accessible to earlier photometric or spectroscopic surveys.
\citet{Rose2026} presented a membership analysis of IC~2395
within the inner $1\degr$ of the cluster centre, establishing the proper
motion, parallax, and photometric selection criteria for cluster membership. In
this work, I apply the same kinematic methodology to an extended search radius
of $2.0\degr$, yielding 173 candidate members and opening the outer cluster
population to infrared disc analysis for the first time using the AllWISE mid-infrared survey.

\section{Data and Membership}
\label{sec:data}
\subsection{Membership catalogue}

The cluster parameters adopted here --- age $\sim$9\,Myr, distance $\sim$750\,pc, and $A_V = 0.6$\,mag --- follow \citet{Rose2026}, where the derivation and associated uncertainties are discussed in detail. Literature values for the extinction span 0.21 to 0.76\,mag across different studies, with a differential extinction of up to 1.42\,mag reported across the cluster field \citep{HuntReffert2023, Perren2023}. This spread is consistent with the scatter observed in the CMD and confirms that a global mean extinction is an approximation across the full 2\degr\ field. The sensitivity of derived stellar masses to the assumed age is addressed in Section~\ref{sec:masses}, where a $\pm$2\,Myr age uncertainty is shown to produce a mean mass shift of $\pm$0.04\,M$_\odot$.

The kinematic membership selection applied here follows \citet{Rose2026}. The Gaia DR3 source list was queried within 2.0\degr\ of the cluster centre (RA 130.425\degr, Dec $-$48.1\degr, J2000), returning 14,832 sources to $G < 19$. Three sequential hard cuts were then applied, as summarised in Table~\ref{tab:selection}: (1) an astrometric quality filter retaining sources with RUWE $< 1.4$ (11,245 sources); (2) a parallax filter $1.2 < \varpi < 1.5$\,mas (412 sources); and (3) a proper motion filter within the cluster proper motion locus, yielding the final 173 members. The proper motion selection applies hard box cuts of $\mu_{\alpha}\cos\delta$ between $-5.7$ and $-4.7$\,mas\,yr$^{-1}$ and $\mu_{\delta}$ between $+3.2$ and $+4.8$\,mas\,yr$^{-1}$, enclosing the cluster locus at $(\mu_{\alpha}\cos\delta,\,\mu_{\delta}) = (-5.2, +4.0)$\,mas\,yr$^{-1}$ with a half-width of 0.5\,mas\,yr$^{-1}$ in each dimension. This is a hard cut, not a probabilistic membership determination; all sources within the proper motion box and satisfying the parallax and quality criteria are included, with no membership probability threshold applied.

The parallax box ($1.2$--$1.5$\,mas, width $0.3$\,mas) is a pragmatic membership filter comfortably bracketing the cluster parallax ($\sim$1.33\,mas; \citealt{Rose2026}) and is not a physically motivated boundary derived from the cluster's line-of-sight depth. The Lindegren et al.\ (2021) parallax zero-point correction \citep{Lindegren2021} was not applied, as no scientific result in this paper depends on individual corrected parallax distances: the cluster distance and age are adopted from \citet{Rose2026}, and all disc fraction calculations are based on photometry and proper motions. The binding membership constraint is the proper motion cut, which is entirely internal to DR3.

\begin{figure*}
  \centering
  \includegraphics[width=\textwidth]{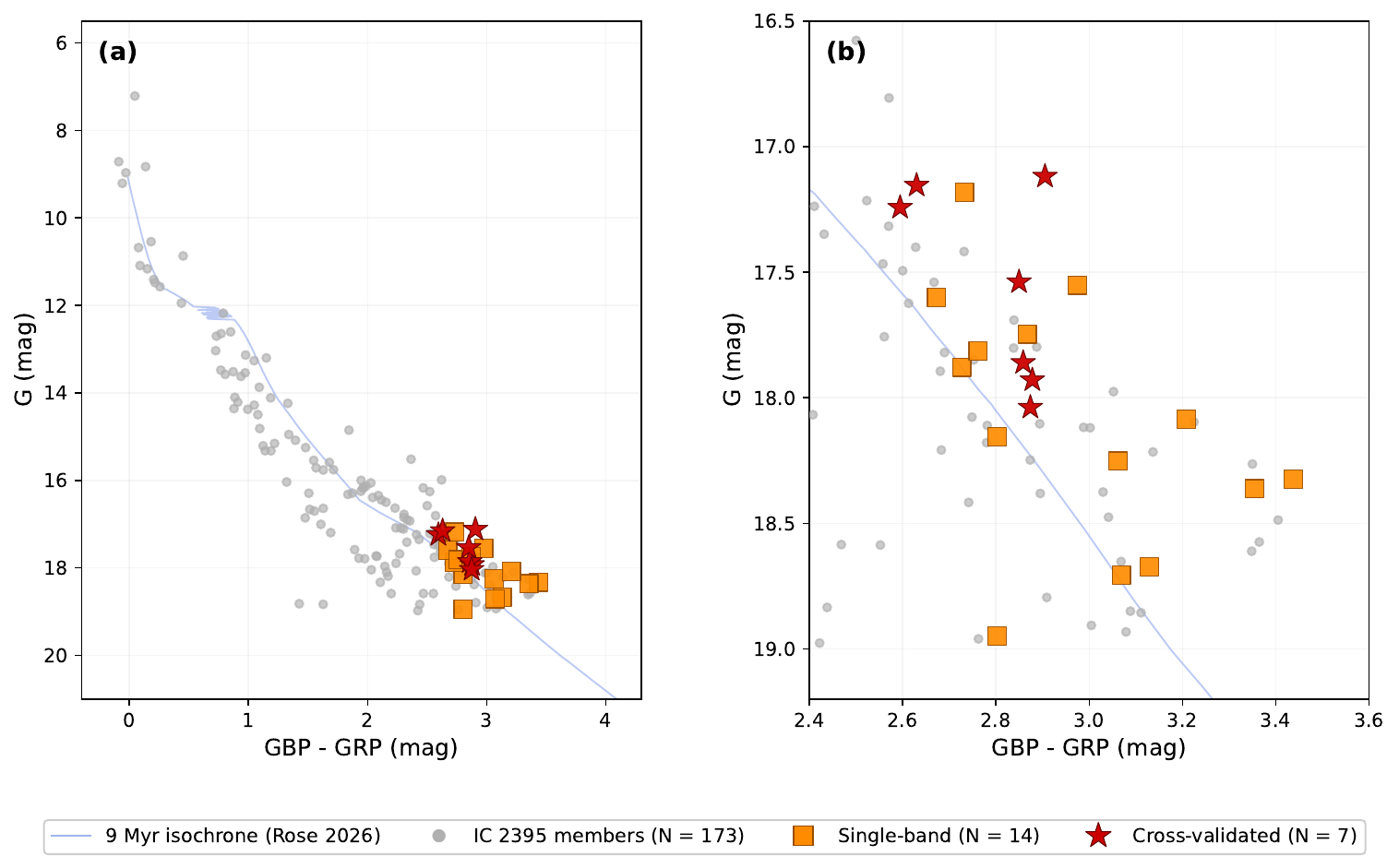}
  \caption{Colour-magnitude diagram of the 173 \object{IC\,2395} kinematic members. Panel (a) shows the full CMD; panel (b) shows the zoomed region containing the 21 disc candidates. 
Cross-validated (red stars) and Single-band (orange squares) disc candidates are shown by tier. The faint blue line shows the 9\,Myr PARSEC v2.0 isochrone \citep{Costa2019, Nguyen2022, Nguyen2025} for reference, adopting $A_V = 0.6$ mag and a distance modulus of 9.38\,mag ($\sim$750\,pc) following \citet{Rose2026}. The scatter about the sequence reflects a combination of photometric uncertainties increasing toward the faint limit ($G \sim 19$), differential extinction across the 2\degr\ field, and a residual field contamination fraction expected from the kinematic selection at large angular separations. An equivalent scatter is present in the 
independently derived membership catalogues of \citet{CantatGaudin2020}, \citet{HuntReffert2023}, and \citet{Perren2023}, confirming it as a genuine feature of the IC\,2395 stellar population rather than an artefact of the kinematic selection. Stars displaced above the main sequence 
by up to $\sim$0.75\,mag are consistent with unresolved binaries; equal-mass pairs produce the maximum displacement of $\sim$0.75\,mag ($\Delta m = 2.5 \log 2$), with lower mass-ratio systems falling between the single-star sequence and this upper envelope.}
\label{fig:cmd_9Myr}
\end{figure*}

The proper motion and parallax filters applied at all separations ensure that field contamination remains controlled across the full 2\degr\ search radius, with no relaxation of membership criteria in the outer field. As a further consistency check, the 5.2$\times$ enhancement of disc candidates above a matched control field (Section~\ref{sec:control}) confirms that the kinematic selection is not dominated by field contaminants. The colour-magnitude diagram of the selected members (Figure~\ref{fig:cmd_9Myr}) provides an independent photometric consistency check: the 173 members trace the 9\,Myr PARSEC isochrone well, confirming that the kinematic selection has returned a genuine cluster population rather than a field-contaminated sample. The scatter about the sequence reflects a combination of photometric uncertainties, differential extinction, and residual field contamination expected from the selection criteria at large angular separations.

The reliability of the kinematic selection was assessed by cross-matching the 173-member catalogue against three independent literature membership catalogues for \object{IC\,2395}: \citet{CantatGaudin2020}, \citet{HuntReffert2023}, and \citet{Perren2023}. Of the 173 members, 99 (57\%) are independently recovered by \citet{HuntReffert2023}, 25 (14\%) by \citet{Perren2023}, and 16 (9\%) by \citet{CantatGaudin2020}. The low overlap with \citet{Perren2023} reflects their more restricted search radius: 21/25 recovered members (75\%) lie within 0.5\degr\ of the cluster centre, dropping to 4/53 (8\%) at 0.5--1.0\degr\ and zero beyond 1.0\degr. The low overlap with \citet{CantatGaudin2020} reflects methodological differences rather than a search radius limit: their UPMASK probabilistic algorithm applied to Gaia DR2 selects a different inner-cluster population from the hard kinematic cuts on DR3 applied here, and the two input datasets (DR2 versus DR3) are not directly comparable at the source\_id level. Critically, of the 74 members unique to this work (unrecovered by \citealt{HuntReffert2023}), 73/74 (99\%) lie beyond 0.5\degr\ from the cluster centre, and 19 of the 21 disc candidates lie beyond 0.5\degr\ (separations 0.23--1.95\degr). These unique members and disc candidates are concentrated precisely in the region where comparison catalogues have low or zero coverage by construction of their search radii, and do not represent a membership disagreement. The strong overlap with \citet{HuntReffert2023} in the inner cluster confirms that the kinematic selection recovers genuine members, while the unique outer population represents new detections enabled by the extended 2\degr\ search radius.

The colour-magnitude diagram of the 173 kinematic members is shown in Figure~\ref{fig:cmd_9Myr}, with the 9\,Myr PARSEC isochrone overlaid.

\begin{table}[!ht]
\caption{Summary of Gaia DR3 kinematic member selection for IC~2395.}
\label{tab:selection}
\centering
\resizebox{\columnwidth}{!}{
\begin{tabular}{lrr}
\hline\hline
\noalign{\vskip 2pt}
Selection Step & Criterion & $N$ \\
\noalign{\vskip 2pt}
\hline
\noalign{\vskip 2pt}
Initial Query & $r < 2.0\degr, G < 19$ & 14,832 \\
Astrometric Quality & RUWE $< 1.4$ & 11,245 \\
Parallax Filter & $1.2 < \varpi < 1.5$\,mas & 412 \\
Proper Motion & within PM box ($\pm$0.5\,mas\,yr$^{-1}$) & 173 \\
\hline
\end{tabular}
}
\tablecomments{The final catalogue of 173 members forms the basis for the infrared disc census.}
\end{table}

\subsection{Photometric data}

AllWISE photometry \citep{AllWISE2019} was cross-matched against the 173 kinematic members using a matching radius of 3\arcsec. 2MASS photometry \citep{Skrutskie_2006} was cross-matched using a matching radius of 2\arcsec. All 152 AllWISE-matched sources have W1 and W2 photometric quality flags of A, indicating signal-to-noise ratio greater than 10 in both bands. W3 and W4 quality flags range from A to U across the matched sample, reflecting the varying reliability of photometry at longer wavelengths in this field.

Of the 173 kinematic members, 152 have reliable AllWISE counterparts within the 3\arcsec\ matching radius. The remaining 21 members have no AllWISE detection, consistent with their faint magnitudes approaching the AllWISE sensitivity limit at the distance of \object{IC 2395}. 

Gaia DR3 astrometric positions are reported at epoch J2016.0; the AllWISE source catalogue positions correspond to a mean epoch of approximately J2010.5. Over the 5.5\,yr baseline, the typical IC\,2395 proper motion of $\sim$6\,mas\,yr$^{-1}$ (total) produces a positional displacement of $\sim$33\,mas. This is well within both the 3\arcsec\ matching radius and the AllWISE astrometric frame uncertainty of $\sim$50--75\,mas (the AllWISE frame is tied to 2MASS, itself tied to the ICRF at $\sim$50\,mas), and epoch propagation was therefore not performed prior to cross-matching. To confirm that the false match rate at 3\arcsec\ is negligible, an offset-and-rematch test was performed: the Gaia positions were displaced by 10\arcsec\ in RA and rematched against 246,956 AllWISE sources within a $4\degr \times 4\degr$ bounding box centred on \object{IC\,2395}. This returns 0 spurious matches from 173 trials, confirming that chance coincidences at 3\arcsec\ are negligible at this Galactic latitude. The W1 and W2 photometric quality flags of A indicate signal-to-noise ratio greater than 10 in both bands; they are a measure of photometric reliability and do not reflect positional accuracy of the cross-match.

Disc fraction estimates are computed against both the full 173-member catalogue and the 152-member AllWISE-matched sample where appropriate.

\section{Disc Identification Method}
\label{sec:method}
\subsection[Primary diagnostic: W1-W2 excess]{Primary diagnostic: W1$-$W2 excess}

The W1$-$W2 colour (3.4$-$4.6\,$\mu$m) was adopted as the primary disc diagnostic. At these wavelengths, excess emission above the stellar photosphere traces hot inner-disc dust at sub-AU scales, probing the primordial disc regime \citep{Dullemond2010, Muzerolle2003}. A threshold of W1$-$W2 $> 0.10$\,mag was applied, based on the observed scatter in the control field stellar locus (Section~\ref{sec:control}). 

W1$-$W2 is a differential colour within the AllWISE photometric system, observed simultaneously with the same instrument. Any diffuse nebular emission that systematically elevates W1 flux will also affect W2 in a correlated manner, rendering the colour difference less susceptible to field-wide additive contamination than cross-instrument colours such as K$-$W1. Its reliability in this field is confirmed empirically by the control field analysis described in Section~\ref{sec:control}.

\subsection[Exclusion of W3-W4 and K-W1 diagnostics]{Exclusion of W3$-$W4 and K$-$W1 diagnostics}

\begin{figure*}
  \centering
  \includegraphics[width=0.75\textwidth, height=0.75\textheight, keepaspectratio]{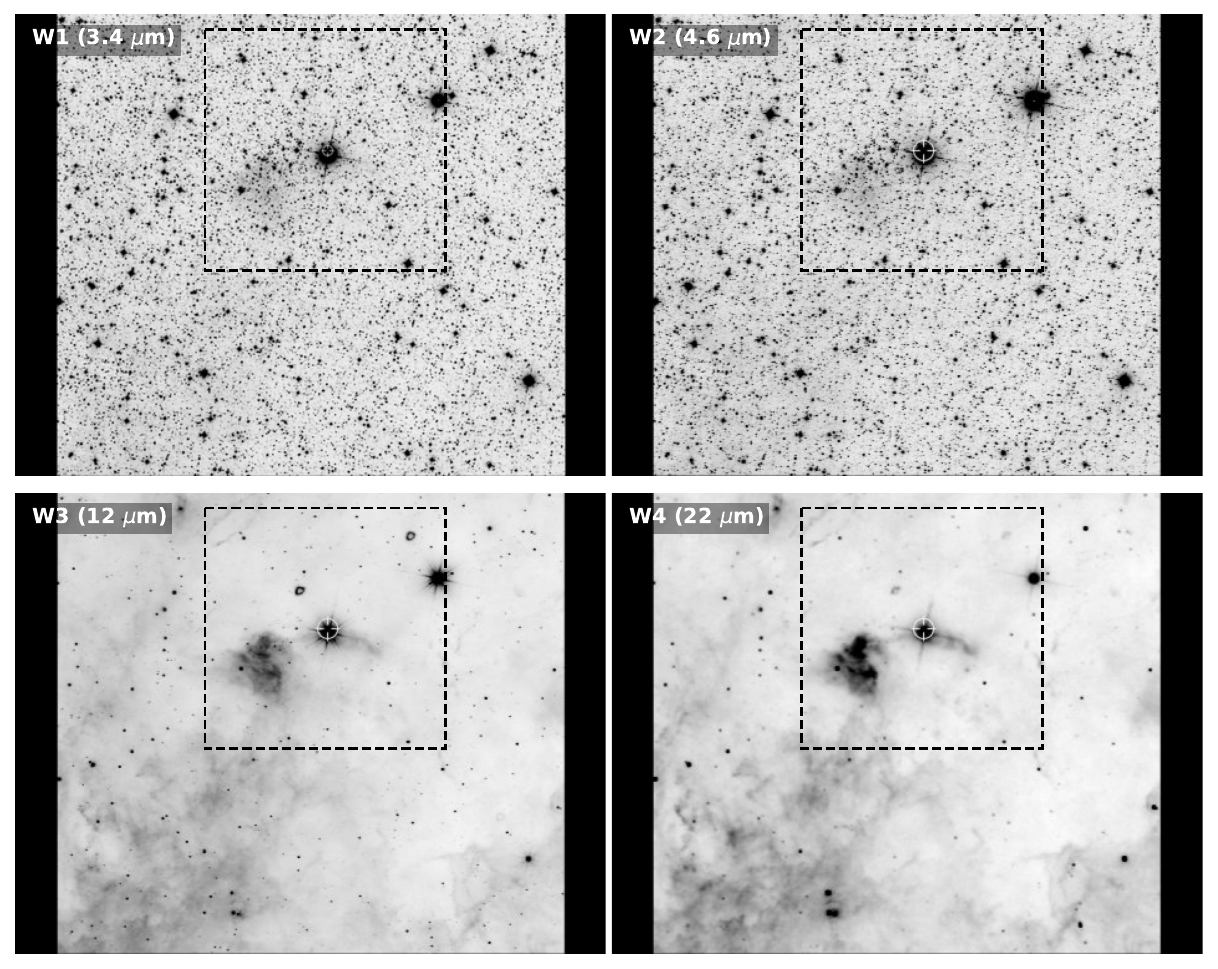}
  \caption{WISE images of the \object{IC 2395} field. The W1 (3.4\,$\mu$m) and W2 
  (4.6\,$\mu$m) panels show a clean field suitable for point-source photometry. 
  The W3 (12\,$\mu$m) and W4 (22\,$\mu$m) panels reveal pervasive diffuse 
  nebular emission from the surrounding \object{Vela} star-forming region, confirming 
  these bands are unsuitable for disc diagnostics in this field.}
  \label{wise_images}
\end{figure*}

The WISE W3 (12\,$\mu$m) and W4 (22\,$\mu$m) bands are excluded from all disc analysis in this work. \object{IC 2395} is embedded in diffuse nebular emission from the surrounding \object{Vela} star-forming region. The large WISE beam at these wavelengths ($\sim$6\arcsec\ at W3, $\sim$12\arcsec\ at W4) is highly susceptible to extended nebular contamination, artificially inflating fluxes for faint point sources. Figure~\ref{wise_images} illustrates this contamination directly: while the \object{IC 2395} field appears clean in W1 and W2, the W3 and W4 images reveal pervasive diffuse emission across the cluster field. Inspection of W3$-$W4 colours during analysis revealed excesses of up to 3$-$4\,mag inconsistent with any stellar or disc model, confirming nebular emission as the dominant signal at these wavelengths.

The K$-$W1 colour (2MASS $K$ at 2.2\,$\mu$m\ versus AllWISE W1 at 3.4\,$\mu$m) was evaluated as a potential disc diagnostic and rejected on statistical grounds. Among the 45 kinematic members with no excess in any other disc diagnostic (W1$-$W2 $\leq 0.10$, W2$-$W3 $\leq 0.30$, H$-$K $\leq 0.20$), the mean K$-$W1 is 0.115\,mag --- significantly above the photospheric value expected for unreddened K/M pre-main-sequence stars ($t = 5.72$, $p = 8.82 \times 10^{-7}$). The false positive rate at the standard 0.20\,mag threshold is 18\% (8/45 confirmed disc-free stars). Disc-bearing and disc-free populations are statistically indistinguishable in K$-$W1 ($t = 1.56$, $p = 0.124$). The diagnostic has no discriminatory power in this field.

This systematic offset most likely reflects PAH emission at 3.3\,$\mu$m\ contaminating the WISE W1 band from the surrounding nebula, and/or diffuse $K$-band nebular emission inflating 2MASS photometry for faint point sources. Unlike W1$-$W2, K$-$W1 is a cross-instrument, cross-mission colour observed over a decade apart with different telescopes and sky subtraction strategies, making it vulnerable to intrinsic YSO variability and field-wide systematic offsets. No ground-based $L$-band catalogue exists for \object{IC 2395} that would enable a valid Haisch-style analysis; the \citet{Balog2016} Spitzer 3.6\,$\mu$m data covers only the central 44\arcmin\ $\times$ 44\arcmin\ footprint, with the wider cluster field surveyed here for the first time.

\subsection{Secondary diagnostics}

W2$-$W3 and H$-$K colours were examined as potential secondary diagnostics but are not used as classification criteria in this work. The W3 band (12\,$\mu$m) is subject to pervasive PAH emission contamination from the surrounding Vela star-forming region, which artificially brightens W3 fluxes field-wide and produces spurious W2$-$W3 excesses independent of any disc emission. This is demonstrated directly by the control field analysis (Section~\ref{sec:control}), in which 75\% of field stars satisfy W2$-$W3 $> 0.30$\,mag despite having no W1$-$W2 excess, and is consistent with the findings of \citet{Koenig2014}, who show that AllWISE W3 fake source rates exceed 75\% in nebular Galactic Plane environments even after photometric quality filtering. No archival data at intermediate wavelengths --- including Spitzer IRAC, which is limited to the central 44\arcmin\ $\times$ 44\arcmin\ footprint of \citet{Balog2016} --- are available to substitute for W3 across 
the wider cluster field. W1$-$W2 therefore remains the sole reliable disc diagnostic for the outer \object{IC\,2395} population surveyed here, and disc candidates are classified into two tiers based on this criterion alone, as described in Section~\ref{sec:classification}.

\begin{figure}
  \centering
  \includegraphics[width=\columnwidth,keepaspectratio]{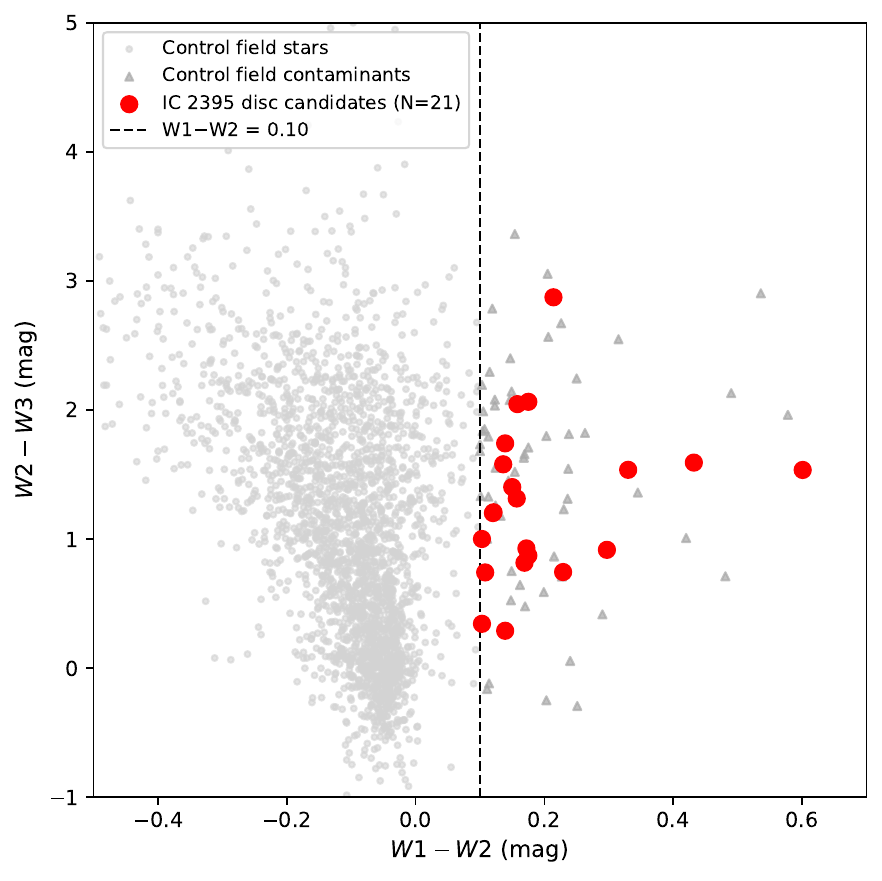}
  \caption{WISE colour-colour diagram for \object{IC 2395} kinematic members. 
  The W1$-$W2 excess criterion (vertical dashed line, W1$-$W2 $> 0.10$\,mag) is shown as the primary disc diagnostic. Control field stars (light grey) and control field contaminants (dark grey triangles) are shown for comparison, illustrating the field-wide W2$-$W3 scatter driven by PAH emission contamination from the surrounding Vela star-forming region. All 21 \object{IC\,2395} 
  disc candidates (red circles) are shown without tier distinction, as W2$-$W3 is not used as a classification criterion in this field 
  \citep{Koenig2014}.}
  \label{wise_ccd}
\end{figure}

\section{Results}
\label{sec:results}

\subsection{Control field analysis}
\label{sec:control}

To assess the contamination rate of the disc selection criteria, two control fields were constructed at the same Galactic latitude as \object{IC 2395} ($b = -3.5\degr$), offset by $\pm10\degr$ in Galactic longitude. The same parallax, RUWE, and magnitude cuts applied to the \object{IC 2395} membership were applied to both control fields, but without proper motion filtering, to produce a representative sample of field stars at the same Galactic depth and subject to the same diffuse nebular environment. Both control fields were triple-matched against AllWISE and 2MASS, yielding a combined sample of 2442 stars.

The fraction of control field stars satisfying the W1$-$W2 $> 0.10$\,mag threshold is 2.4\% (58/2442). Among \object{IC 2395} kinematic members, 21 of the 152 AllWISE-matched stars satisfy this threshold, a fraction of 13.8\% (the remaining 21 members lack AllWISE detections due to their faint magnitudes approaching the survey sensitivity limit; disc fractions are computed against the full 173-member catalogue throughout). The enhancement factor of \object{IC\,2395} disc candidates above the control field background is 5.2$\times$ at the 
W1$-$W2 $> 0.10$\,mag criterion. This enhancement above a field matched for Galactic latitude, distance, and nebular environment confirms that the disc candidates represent a genuine population of cluster members with infrared excess, rather than field contaminants. Applying the control field contamination rate of 2.4\% (58/2442; see above) to the 152 AllWISE-matched members 
yields an expectation of $\sim$4 false positives among the 21 candidates; the observed 5.2$\times$ enhancement above this background confirms that the majority represent a genuine disc-bearing population. The control field stellar locus is shown alongside the \object{IC\,2395} disc candidates in Figure~\ref{wise_ccd}.

The W1$-$W2 $> 0.10$ mag threshold was chosen empirically based on the observed dispersion of the control field stellar locus, ensuring that any systematic offsets -- whether from subtle photospheric reddening in late-type stars or field-wide PAH emission at 3.3\,$\mu$m -- are absorbed into the statistical baseline. The 5.2$\times$ enhancement above this baseline constitutes the primary evidence that the disc candidates represent a genuine cluster population. The control fields are not spatially restricted to the 2\degr\ search radius but are drawn from the same Galactic latitude ($b = -3.5\degr$) and depth (same parallax and magnitude 
selection), placing them in the same diffuse nebular environment. Differential reddening is negligible: $E(W1-W2) < 0.01$ mag for $A_V = 0.6$ mag \citep{Wang2019}, consistent across both cluster and control fields. The control field population skews slightly earlier in 
spectral type than the kinematically confirmed cluster members; any intrinsic W1$-$W2 reddening of late-type photospheres relative to earlier-type stars is therefore not captured in 
the baseline, but this acts conservatively --- making the threshold marginally harder to exceed for the reddest cluster members rather than inflating the candidate count.

\begin{figure}
  \centering
  \includegraphics[width=\columnwidth,]{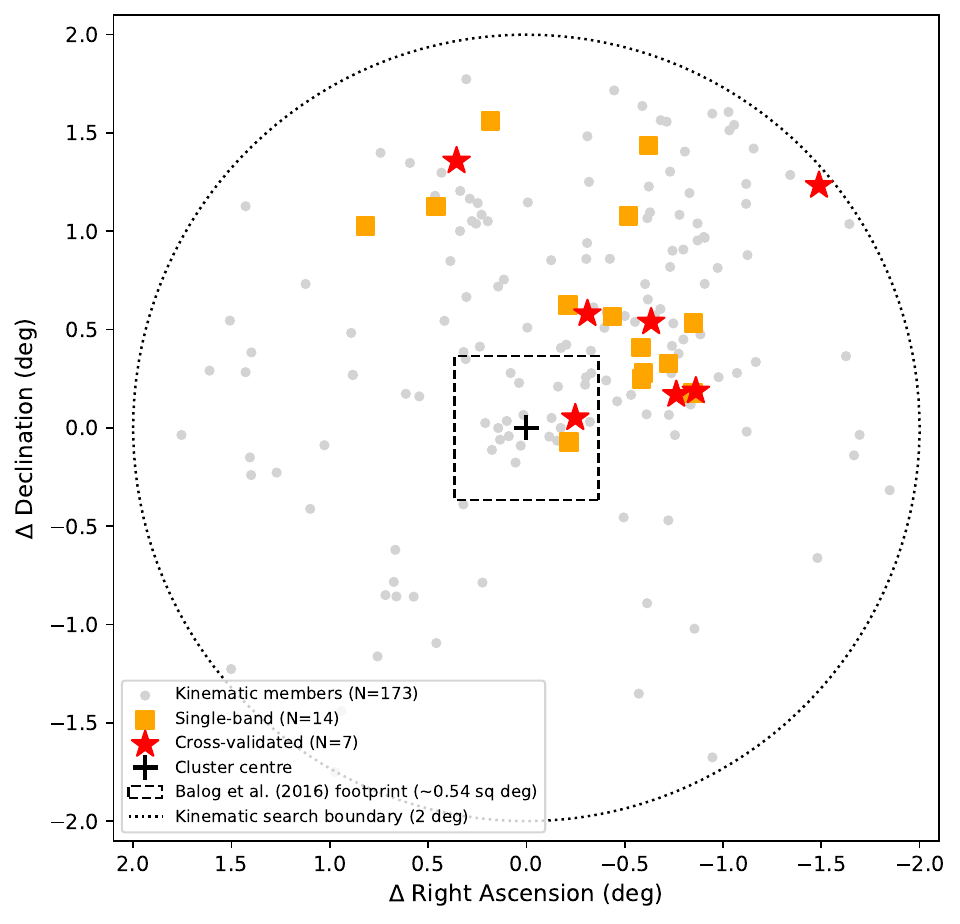}
  \caption{Spatial distribution of \object{IC 2395} kinematic members and disc candidates. 
  Grey points show all 173 kinematic members. Cross-validated (red stars) and Single-band (orange squares) disc candidates are shown by tier. The dashed square marks the 44\arcmin\ $\times$ 44\arcmin\ Spitzer survey footprint of \citet{Balog2016}, within which only 2 of the 21 disc candidates fall. The dotted circle marks the 2\degr\ kinematic search boundary. Right 
  Ascension increases to the left following the astronomical convention.}
  \label{fig:spatial}
\end{figure}

\subsection{Classification scheme}
\label{sec:classification}

Disc candidates were classified into two tiers based on the availability of independent corroborating evidence. The tier system is intended as a confidence metric rather than a disc evolutionary classification.

Cross-validated disc candidates ($N = 7$) are required to satisfy W1$-$W2 $> 0.10$\,mag and to be independently classified as young stellar objects by the Gaia DR3 variability classifier. The Gaia 
YSO classification is based on multi-epoch photometric variability in the $G$, $G_\mathrm{BP}$, and $G_\mathrm{RP}$ bands over the 2014--2022 baseline, entirely independent of the AllWISE photometry used for the primary disc diagnostic. This independence makes the Cross-validated tier the most robust: infrared excess and photometric variability evidence for youth are satisfied simultaneously. Cross-validated candidates are retained even where their W1$-$W2 excess falls below $3\sigma$ photometric significance, because the Gaia variability classification constitutes an orthogonal line of evidence for the presence of an actively accreting disc that is entirely independent of the AllWISE photometry.

Single-band disc candidates ($N = 14$) satisfy W1$-$W2 $> 0.10$\,mag but lack an independent Gaia YSO variability classification. As discussed in Section~\ref{sec:method}, W2$-$W3 and H$-$K colours cannot be used as reliable secondary diagnostics in this field due to pervasive PAH emission contamination of the W3 band from the surrounding Vela star-forming region \citep{Koenig2014}. The W1$-$W2 excess criterion is therefore the sole basis for their 
classification as disc candidates, and follow-up observations are required to confirm their disc-bearing status.

 It is noted that W1$-$W2 excess at 3.4 - $4.6\,\mu$m traces only the innermost dust, and cannot distinguish between primordial, transitional, and debris disc material. Single-band candidates should therefore be understood as infrared excess sources with properties consistent with inner-disc emission, pending longer-wavelength confirmation.

 No epoch photometry is available in Gaia DR3 for any of the 14 Single-band candidates, and none were flagged as variable by the Gaia DR3 variability pipeline. The photometric significance of the W1$-$W2 excess above the 0.10\,mag threshold, defined as $\Sigma = (W1-W2 - 0.10)/\sqrt{\sigma_{W1}^{2} + \sigma_{W2}^{2}}$, is reported for each candidate in Table~\ref{tab:full_appendix}. For the Single-band tier, $\Sigma$ ranges from $0.07\sigma$ to $7.83\sigma$ (mean $2.1\sigma$), with several candidates near the detection limit. This solidifies the need for follow-up observations.

All 21 disc candidates satisfy a renormalised unit weight error below 1.4 and have astrometric solutions consistent with single stars in Gaia DR3, excluding sources where unresolved binarity could mimic infrared excess through flux contamination.

\subsection{Spatial distribution}
\label{sec:spatial}

The spatial distribution of disc candidates is shown in Figure~\ref{fig:spatial}. All 21 candidates lie at angular separations of 0.23\degr--1.95\degr\ from the cluster centre (RA 130.425\degr, Dec $-$48.1\degr, J2000). Nineteen of the 21 candidates lie beyond the 44\arcmin\ $\times$ 44\arcmin\ Spitzer survey footprint of \citet{Balog2016}, confirming that this population was entirely inaccessible to prior infrared disc surveys of \object{IC 2395}. The remaining two candidates near or within the \citet{Balog2016} footprint were not recovered by that survey.

No strong radial dependence of disc fraction with cluster-centric separation is detected within the sample, though the small number of candidates precludes a statistically robust test of disc survival as a function of projected distance from the cluster centre.

Within 0.5\degr\ of the cluster centre, 2 of 28 members are disc candidates (7.1\%); beyond 0.5\degr, 19 of 145 members are disc candidates (13.1\%). The small number of candidates in the inner bin precludes a statistically robust comparison between the two regions.

\subsection{Disc fraction}
\label{sec:discfraction}

Three complementary disc fraction estimates are derived, each probing a different confidence threshold.

The \textit{photometric $3\sigma$ disc fraction} applies a per-star significance criterion $\Sigma = (W1-W2 - 0.10)/\sqrt{\sigma_{W1}^{2} + \sigma_{W2}^{2}} \geq 3$ to the 21 disc candidates. Five of 21 candidates survive this cut (2 Cross-validated, 3 Single-band), giving a raw fraction of $5/173 = 2.9 \pm 1.3$\%. To assess the false positive rate at $3\sigma$, per-star excess significance was computed for the full control field sample of 2436 stars with valid AllWISE uncertainties: 14 satisfy $\Sigma \geq 3$ ($0.575\%$). The expected number of $3\sigma$ false positives among the 152 AllWISE-matched members is therefore $0.575\% \times 152 = 0.87$ stars, giving a background-subtracted $3\sigma$ fraction of $\sim$$2.4 \pm 1.3$\%. The enhancement of \object{IC\,2395} disc candidates above the control field at $3\sigma$ is 5.7$\times$.

The \textit{Cross-validated disc fraction} is $7/173 = 4.0 \pm 1.5$\%. These seven stars satisfy W1$-$W2 $> 0.10$\,mag and are independently classified as YSOs by the Gaia DR3 variability classifier based on multi-epoch photometric variability over the 2014--2022 baseline (Section~\ref{sec:variability}). The Gaia classification is entirely independent of the AllWISE photometry. As discussed in Section~\ref{sec:classification}, this independence permits Cross-validated candidates to be retained as robust disc candidates even where their W1$-$W2 excess falls below $3\sigma$: the variability confirmation constitutes an orthogonal line of evidence for disc presence. Background subtraction is not applied to this tier, as the individual variability confirmation supersedes the need for a statistical correction. The Cross-validated fraction ($4.0 \pm 1.5$\%) is adopted as the headline result.

The \textit{full-sample upper bound} is $21/173 = 12.1 \pm 2.5$\%, based on all candidates satisfying W1$-$W2 $> 0.10$\,mag. As W1$-$W2 traces only warm inner-disc emission and cannot distinguish between primordial, transitional, and debris disc material, this figure should be interpreted as a soft upper limit; any residual field contamination is expected to reside preferentially within the Single-band tier. Applying the control field contamination rate of 2.4\% (58/2442) to the 152 AllWISE-matched members yields an expectation of $\sim$3.6 false positives among the 21 candidates; a background-subtracted upper bound of $\sim$10.1\% is therefore derived. The uncertainty on this figure is computed by combining the binomial uncertainty on the 21 detections with the Poisson uncertainty on the expected false positive count ($\sqrt{58}/2442 \times 152 = 0.47$ stars), giving a total uncertainty of $\pm 2.5$\%.

The convergence of the photometric $3\sigma$ estimate ($2.9\%$) and the Cross-validated estimate ($4.0\%$) --- derived by entirely independent methods --- supports a secure disc fraction in the range $2.9$--$4.0$\% for \object{IC\,2395}. All figures are calculated against the full 173-member kinematic catalogue to ensure a conservative estimate that accounts for members below the AllWISE detection limit.

\setlength{\tabcolsep}{18pt} 
\begin{table*}[ht]
\caption{Variability characteristics of Tier 1 (Cross-validated) disc candidates.}
\label{tab:variability}
\centering
\begin{tabular}{lccc}
\hline\hline
Gaia Source ID & $G$ (mag) & Variability Class & Physical Interpretation \\
\hline
5329380722948000640 & 17.93 & Dramatic dipper & High-inclination occultation \\
5329301253185618432 & 17.54 & Dramatic burster & Unsteady accretion episode \\
5329277029570154368 & 17.12 & Mixed / episodic & Complex inner-disc structure \\
5329761055204021888 & 18.04 & Single outburst & Accretion burst/flare \\
5329217071827655808 & 17.24 & Stable / weak & Low-rate steady accretion \\
5329266476819288448 & 17.86 & Burster & Magnetospheric hotspots \\
5521710874599156480 & 17.15 & Episodic burster & Modulated mass transfer \\
\hline
\end{tabular}
\tablecomments{Variability classifications are derived from Gaia DR3 epoch photometry and represent independent evidence for the presence of structured circumstellar material.}
\end{table*}
\setlength{\tabcolsep}{6pt} 

\section{Stellar masses}
\label{sec:masses}

Stellar masses for the 21 disc candidates were estimated by interpolating along the PARSEC v2.0 isochrone \citep{Costa2019, Nguyen2022, Nguyen2025} at a cluster age of $\sim$9\,Myr and solar metallicity ($Z = 0.0152$). A distance modulus of 9.38\,mag and extinction $A_V = 0.6$\,mag were adopted, following \citet{Rose2026}. The Gaia $G$-band magnitude was chosen as the primary interpolation axis; while accretion veiling can affect broad-band photometry, the $G$ filter is significantly less sensitive to blue-excess veiling than $G_\mathrm{BP}$ \citep{Manara2013, Nguyen2022}.

The full sample of 21 candidates falls within a narrow mass range of 0.31--0.68\,M$_\odot$, with both mean and median masses of $\sim$0.50\,M$_\odot$. The mass range of disc candidates reflects the kinematic completeness limit of the Gaia DR3 membership catalogue at $G \sim 19$\,mag at the distance of IC\,2395, below which astrometric uncertainties preclude reliable membership confirmation under the strict proper motion and parallax cuts applied here.  This concentration among K- and early M-type stars is physically consistent with the mass-dependent dissipation of primordial discs, where higher-mass stars in \object{IC 2395} ($\gtrsim$1\,M$_\odot$) have already transitioned to the debris disc or disc-less phase by $\sim$9\,Myr \citep{ribas2015}.

\begin{figure}[ht]
    \centering
    \includegraphics[width=\columnwidth]{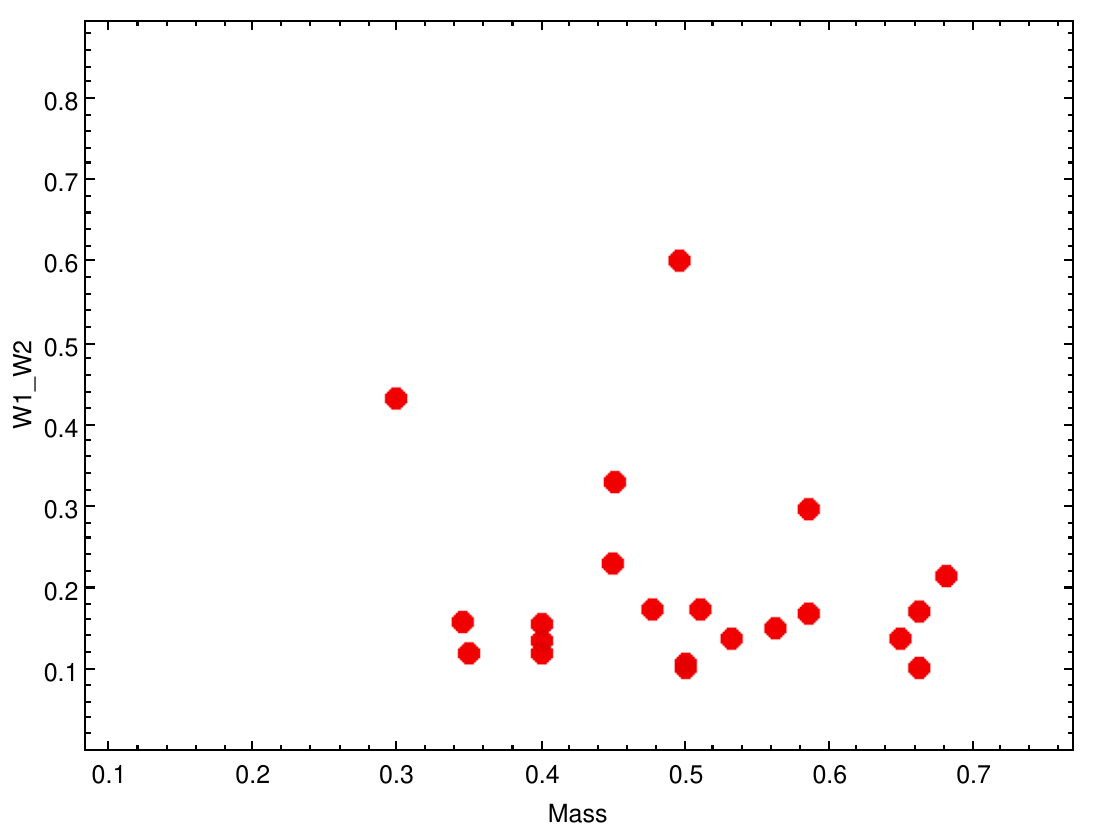}
    \caption{Mid-infrared excess ($W1-W2$) as a function of stellar mass for the 21 disc candidates. No significant correlation is observed ($\rho = +0.07, p = 0.84$), indicating that the strength of the infrared excess at 9\,Myr is independent of host star mass within the K--M dwarf regime.}
    \label{fig:mass}
\end{figure}

The relationship between stellar mass and infrared excess ($W1-W2$) was evaluated using a Spearman rank correlation test, yielding $\rho = +0.07$ ($p = 0.84$), as visualized in Fig.~\ref{fig:mass}. With a sample of only 21 objects spanning a narrow mass range of 0.31--0.68\,M$_\odot$, this test has insufficient statistical power to detect or rule out a weak mass--excess correlation; the result should be understood as uninformative rather than as evidence against a mass dependence. The observed range of W1$-$W2 excess from 0.10 to 0.60\,mag across this narrow mass range is nonetheless consistent with stochastic drivers such as inner-disc geometry or varying accretion rates.

Mass estimates carry an uncertainty  dominated by the assumed $\pm$2\,Myr age spread of the cluster; using 7\,Myr and 11\,Myr isochrones as bounds resulted in a mean mass shift of $\pm$0.04\,M$_\odot$. The impact of near-infrared excess on the $G_\mathrm{RP}$ magnitude was found to be secondary to these model-based uncertainties.

\section{Gaia epoch photometry and YSO variability}
\label{sec:variability}

\begin{figure*}
  \centering
  \includegraphics[width=0.85\textwidth, height=0.85\textheight, keepaspectratio]{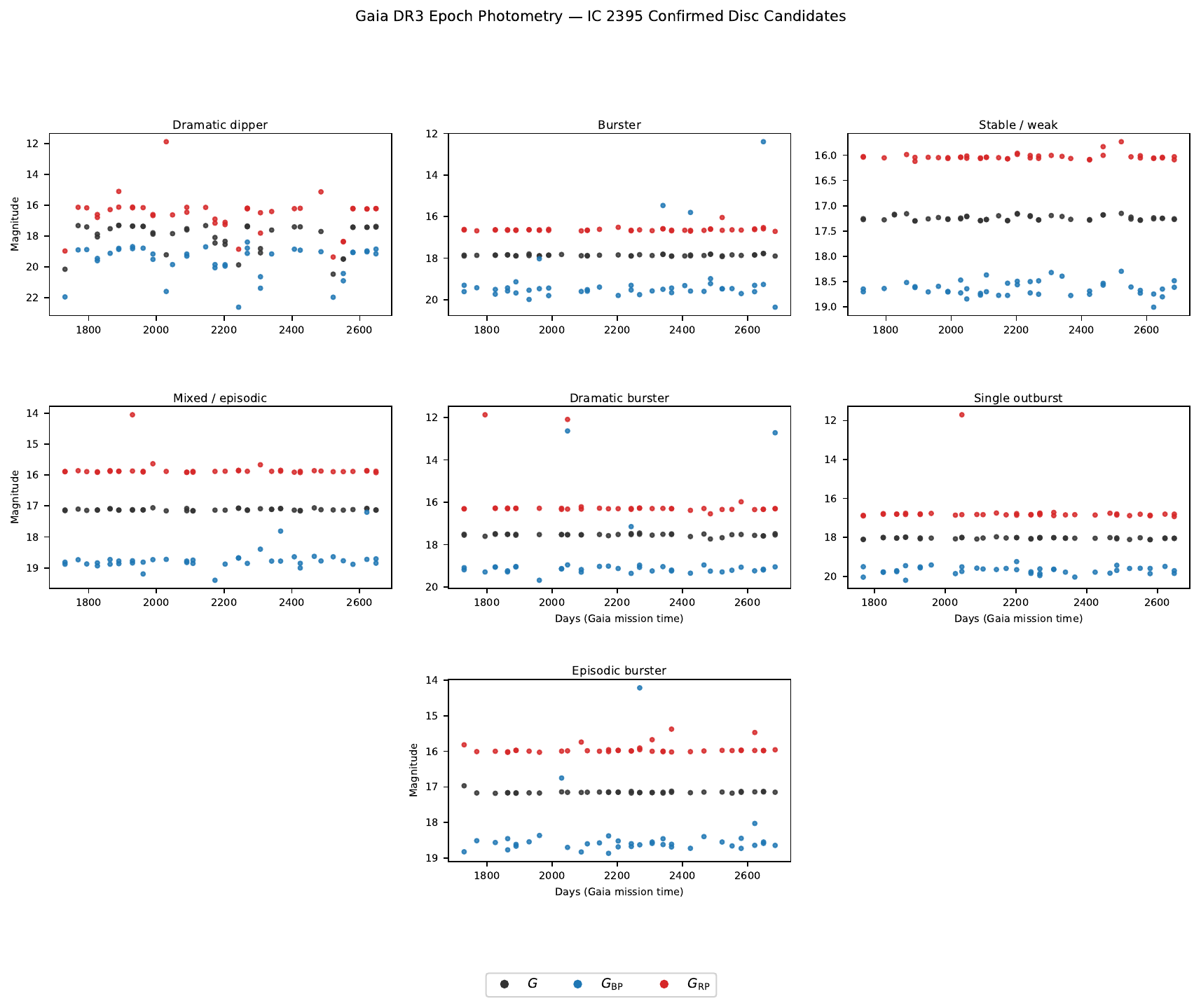}
  \caption{Gaia DR3 epoch photometry for the seven Cross-validated disc candidates 
  in \object{IC 2395}. Individual $G$ (black), $G_\mathrm{BP}$ (blue), and 
  $G_\mathrm{RP}$ (red) band measurements are shown as a function of time 
  over the 2014--2022 Gaia baseline. Diverse variability morphologies are 
  evident, including a dramatic dipper, bursters, a single outburst, mixed 
  and episodic variables, and a stable or weakly variable source, consistent 
  with actively accreting pre-main-sequence stars at $\sim$9\,Myr.}
  \label{fig:IC2395_Fig4_YSO_LightCurves_colour_3x3}
\end{figure*}

The seven Cross-validated disc candidates are independently classified as young stellar objects by the Gaia DR3 variability classifier based on multi-epoch photometric monitoring in the $G$,
$G_\mathrm{BP}$, and $G_\mathrm{RP}$ bands over the 2014--2022 baseline. Gaia epoch photometry for these seven stars is presented in Figure~\ref{fig:IC2395_Fig4_YSO_LightCurves_colour_3x3}, with 270 individual measurements shown for the $G$, $G_\mathrm{BP}$, and $G_\mathrm{RP}$ bands,
revealing a diverse range of variability morphologies characteristic of actively accreting pre-main-sequence stars.

The light curves were classified into descriptive morphological categories based on visual inspection following the framework of \citet{Cody2014}. These classifications and their physical interpretations are summarized in Table~\ref{tab:variability}.

\begin{itemize}
  \item One star shows \textit{dramatic dipper} behaviour, characterised by deep transient dimming events below a stable baseline. Dippers are attributed to occultation of the stellar photosphere by warps, clumps, or accretion columns near the inner disc edge, associated with near-corotation disc structure.

  \item Two stars show \textit{burster} behaviour: one classified as a burster and one as a dramatic burster. Both are characterised by episodic brightenings above a stable baseline, consistent with magnetospheric accretion events delivering material onto the stellar surface in discrete episodes \citep{Stauffer2014}; higher amplitude bursting behaviour has been associated with stronger inner disc infrared excess as measured by W1$-$W2 colour \citep{Cody2017}.

  \item One star shows a \textit{single outburst}, a dramatic brightening isolated to a single epoch across the full Gaia baseline.

  \item One star shows \textit{mixed or episodic} behaviour combining both brightening and dimming events, consistent with a system in which both accretion and occultation contribute to the observed variability.

  \item One star shows \textit{episodic burster} behaviour, characterised by irregular brightening events of varying amplitude across the Gaia baseline, consistent with unsteady magnetospheric accretion.

  \item One star shows \textit{stable or weak} variability, with little detectable modulation across the Gaia baseline, consistent with a disc-bearing star caught in a relatively quiescent accretion state.
\end{itemize}

These variability types are consistent with the range of behaviour observed in classical T\,Tauri stars in younger star-forming regions \citep{Herbst1994, Cody2014}, and their presence in \object{IC 2395} at $\sim$9\,Myr confirms that active accretion is ongoing in at least a subset of the disc candidate population at this age. The detection of a dramatic dipper and three burster-type variables (a burster, a dramatic burster, and an episodic burster) --- typically associated with structured inner disc material at or near the corotation radius --- supports the presence of optically thick primordial discs rather than tenuous debris disc material.

The morphological classifications derived from Gaia epoch photometry should be treated with appropriate caution, however. The sparse and irregular sampling of the Gaia time series --- typically tens of observations over eight years --- means that variability events may fall between observation windows, and stars classified as stable or weakly variable may not necessarily have been observed during active accretion phases. As \citet{Cody2017} note, this remains speculative. Despite the sparse sampling, the high degree of colour-consistency across the $G$, $G_\mathrm{BP}$, and $G_\mathrm{RP}$ bands during flux excursions (as seen in Fig.~\ref{fig:IC2395_Fig4_YSO_LightCurves_colour_3x3}) suggests these are physical events rather than photometric noise.

\section{Discussion}
\label{sec:discussion}

\subsection{Disc fraction in context}
\label{sec:discfrac_context}

The secure disc fraction of \object{IC\,2395}, established by the Cross-validated subsample independently confirmed by Gaia DR3 variability classification, is $4.0 \pm 1.5$\% (7/173). A consistent photometric $3\sigma$ estimate of $2.9 \pm 1.3$\% is obtained from the significance-filtered sample. The full-sample upper bound is $12.1 \pm 2.5$\% (21/173), with a background-subtracted value of $\sim$10.1\%. These values place \object{IC\,2395} at the tail of the primordial disc decay curve. Primordial disc fractions decline approximately exponentially from $\sim$80\% at 1\,Myr to fewer than 10\% by 8--10\,Myr, with a characteristic dissipation timescale of 2--3\,Myr \citep{mamajek2009, fedele2010, richert2018}. At $\sim$9\,Myr, any surviving primordial disc population is therefore scientifically significant, representing the tail of a distribution in which most discs have already dissipated.

Direct comparison with the prior \object{IC 2395} disc census of \citet{Balog2016} is instructive. That study identified 18 protoplanetary disc (6.5\%) and 8 transitional disc (2.9\%) candidates using Spitzer IRAC and MIPS photometry within the 44\arcmin\ $\times$ 44\arcmin\ survey footprint. The present work extends to 2\degr\ from the cluster centre, recovering a disc-bearing population that was entirely outside the Spitzer footprint. The 5.2$\times$ enhancement of disc candidates above the control field background confirms that this outer population represents genuine cluster members with circumstellar discs, rather than field contaminants. This enhancement suggests that the low-mass members of \object{IC 2395} are not just centrally concentrated, but retain their primordial discs regardless of their position within the wider gravitational potential of the cluster. The spatial extent of the disc-bearing population in \object{IC 2395} is shown to be  significantly larger than previously recognised.

Comparison with disc fractions in other clusters of similar age requires care, as different studies employ different diagnostics, membership methods, and spatial coverage. The W1$-$W2 excess used here is most analogous to Spitzer IRAC 8\,$\mu$m excess, tracing hot inner disc dust in the primordial disc regime. Direct numerical comparison with studies that rely on Spitzer MIPS 24\,$\mu$m diagnostics --- which trace cooler debris disc material at larger radii --- is not straightforward; \citet{Smith2011} explicitly confirm that the 42\% 24\,$\mu$m excess fraction they detect in \object{IC 4665} at 27\,Myr arises entirely from debris rather than primordial disc material, illustrating the diagnostic difference \citep{Gorlova2007, Smith2011}. Within the same cluster, \citet{Balog2016} identified 23 debris disc candidates (8.3\%) using Spitzer photometry within the 44\arcmin\ $\times$ 44\arcmin\ survey footprint; as W1$-$W2 cannot distinguish primordial from debris material, this population provides relevant context for interpreting the upper bound of 12.1\%. Within these caveats, the \object{IC\,2395} disc fraction of 12.1\% at $\sim$9\,Myr is broadly consistent with the expected tail of primordial disc decay, and with the disc fraction of \object{Upper Scorpius} ($\sim$11\,Myr) for B--G type stars of less than 10\% \citep{luhman2012}, and with the inner disc decay timescales derived from a recent blind Gaia+WISE survey of 32 clusters spanning 1--100\,Myr \citep{ben2025}. The higher disc fractions seen around lower-mass stars in \object{Upper Scorpius} ($\sim$25\% for spectral types $\geq$M5; \citealt{luhman2012}) are consistent with the mass-dependent survival seen in the \object{IC 2395} disc candidate sample, which is concentrated entirely in the K- and early M-type regime. 

\subsection{Spatial bias in prior surveys and the value of kinematic membership}
\label{sec:spatial_bias}

The key distinguishing result of this work is spatial. The \citet{Balog2016} Spitzer survey covered only the central 44\arcmin\ $\times$ 44\arcmin\ of \object{IC 2395}, leaving the outer cluster field entirely uncharacterised. By applying the kinematic membership criteria of \citet{Rose2026} to a search radius of 2\degr, this work recovers a disc-bearing population that was entirely outside the prior survey footprint. Nineteen of the 21 disc candidates identified here lie beyond the \citet{Balog2016} boundary.

This result illustrates a broader methodological point. Pointed infrared imaging surveys are necessarily limited in angular coverage, and for clusters at intermediate distances such as \object{IC 2395} ($\sim$750\,pc), the accessible field may represent only the dense cluster core. Kinematically selected membership catalogues based on Gaia proper motions and parallaxes remove this spatial bias, enabling disc surveys that extend to the full extent of the kinematic cluster population. The outer disc population recovered here --- enhanced 5.2$\times$ above control field background --- would have been invisible to any prior survey without kinematic pre-selection.

\subsection{Mass dependence and disc survival}
\label{sec:mass_dependence}

All 21 disc candidates fall in the mass range 0.31--0.68\,M$_\odot$, corresponding to K- and early M-type pre-main-sequence stars. No disc candidates are identified among higher-mass members, consistent with the well-established trend of faster disc dissipation around more massive stars \citep{luhman2012, ribas2015}. At $\sim$9\,Myr, the survival of primordial discs is expected predominantly around lower-mass stars, and the \object{IC 2395} disc candidate sample is fully consistent with this expectation.

The absence of a correlation between stellar mass and W1$-$W2 excess within the disc candidate sample (Spearman $\rho = +0.07$, $p = 0.84$) is also physically meaningful. W1$-$W2 traces the instantaneous accretion state of the inner disc rather than the underlying disc mass. The diverse variability morphologies seen in the Cross-validated subsample --- spanning a dramatic dipper, bursters, and episodic variables --- further indicate that the disc candidates are caught at different points in their accretion cycles, rather than representing a mass-ordered population.

\subsection[Accretion at 9 Myr]{Accretion at $\sim$9\,Myr}
\label{sec:accretion}

The detection of actively accreting YSOs in \object{IC 2395} at $\sim$9\,Myr, independently confirmed by Gaia DR3 variability classification, demonstrates that magnetospheric accretion is ongoing in a subset of low-mass cluster members at this age. The presence of a dramatic dipper and three burster-type variables --- variability morphologies directly associated with structured inner disc material at the corotation radius --- supports the interpretation that the Cross-validated disc candidates retain optically thick primordial discs rather than optically thin debris disc material.

The single dramatic dipper in the sample is consistent with the geometric requirement for dipper phenomena: occultation of the stellar photosphere requires the inner disc to be inclined such that disc material --- warps, dust clumps, or accretion columns near corotation --- passes across the line of sight. As \citet{Cody2018} note, dippers favour high disc inclinations. The detection of a single dipper among seven Cross-validated variables is qualitatively consistent with the expectation that only a fraction of disc-bearing stars will be favourably oriented at any given age; the small number of Cross-validated candidates precludes a statistically meaningful comparison with the $\sim$20--30\% dipper fraction reported in larger samples \citep{Cody2014}.

This is consistent with theoretical expectations: at $\sim$9\,Myr, the disc dissipation timescale for K- and M-type stars has not yet elapsed for all members of the population, and a tail of actively accreting systems is predicted. The \object{IC 2395} Cross-validated subsample provides direct observational evidence for this tail at an age where such systems are rare but physically expected.

\section{Conclusions}
\label{sec:conclusions}

I have presented an AllWISE infrared excess survey of the young open cluster \object{IC\,2395}, based on a kinematically selected membership catalogue of 173 stars extending to 2\degr\ from the 
cluster centre. Cross-matching against AllWISE and 2MASS photometry, I identify 21 disc candidates in two confidence tiers, and derive the following principal conclusions:

\begin{enumerate}

\item A secure disc fraction of $4.0 \pm 1.5$\% (7/173) is established by the Cross-validated subsample, independently confirmed by Gaia DR3 YSO variability classification. A consistent photometric $3\sigma$ estimate of $2.9 \pm 1.3$\% is derived from significance-filtered candidates. The full-sample upper bound is $12.1 \pm 2.5$\% (21/173), with a background-subtracted value of $\sim$10.1\%. These values are consistent with the characteristic 2--3\,Myr dissipation timescale \citep{mamajek2009, richert2018} and place \object{IC\,2395} in the expected tail of primordial disc decay at $\sim$9\,Myr.

\item Nineteen of the 21 disc candidates (90\%) lie beyond the 44\arcmin\ $\times$ 44\arcmin\ Spitzer survey footprint of \citet{Balog2016}. This recovery of a spatially extended, low-mass disc population --- enhanced 5.2$\times$ above the control field background --- confirms that prior surveys may have underestimated the spatial extent of disc survival in \object{IC\,2395} due to Spitzer footprint limitations.

\item The disc-bearing population is concentrated entirely in the K- and early M-type regime (0.31--0.68\,M$_\odot$), consistent with faster disc dispersal around higher-mass stars. No significant correlation between stellar mass and W1$-$W2 excess was detected (Spearman $\rho = +0.07$, $p = 0.84$), suggesting that within this mass range, the inner-disc signature is independent of host mass and likely driven by stochastic accretion states or disc geometry.

\item Seven Cross-validated disc candidates are independently classified as young stellar objects by the Gaia DR3 variability classifier, exhibiting a diverse range of variability morphologies including a dramatic dipper, three burster-type variables, and episodic variability, confirming that active magnetospheric accretion is ongoing in \object{IC\,2395} at $\sim$9\,Myr. The detection of a single dramatic dipper is qualitatively consistent with the geometric requirement that dipper phenomena favour high disc inclinations \citep{Cody2018}, though the small Cross-validated sample precludes a statistical comparison.

\item The kinematic membership approach adopted here, extending the search radius well beyond the dense cluster core, recovers a spatially extended disc-bearing population that would be inaccessible to pointed infrared surveys. This demonstrates the value of Gaia-based kinematic pre-selection for unbiased disc fraction measurements in intermediate-distance clusters. 

\end{enumerate}

Follow-up observations could significantly strengthen and extend the results presented here. Near-infrared spectroscopy of the 21 disc candidates would enable direct measurement of accretion 
indicators such as H$\alpha$ emission and UV excess, confirming active accretion and providing quantitative accretion rate estimates for the Cross-validated subsample. High-resolution 
imaging with JWST or ground-based adaptive optics systems would resolve the inner disc structure of the brightest candidates and test the high-inclination hypothesis for the dramatic dipper. 
JWST MIRI photometry at 8--24\,$\mu$m across the full 2\degr\ kinematic field would enable disc diagnostics unaffected by the PAH contamination that renders the WISE W3 band unreliable in 
this field \citep{Koenig2014}, and would extend the disc diagnostic to cooler dust at larger disc radii. Radial velocity monitoring of the 14 Single-band disc candidates would confirm or refute their cluster membership, potentially adding to the confirmed disc-bearing population. Gaia DR4, due in December 2026 and covering 66 months of mission data, will provide an extended photometric baseline that may improve the variability classification for Single-band candidates currently lacking Gaia YSO classifications, potentially promoting some to Cross-validated status.

\section*{Acknowledgements}

This work has made use of data from the European Space Agency (ESA) 
mission \textit{Gaia} (\url{https://www.cosmos.esa.int/gaia}), processed 
by the \textit{Gaia} Data Processing and Analysis Consortium (DPAC, 
\url{https://www.cosmos.esa.int/web/gaia/dpac/consortium}). Funding for 
the DPAC has been provided by national institutions, in particular the 
institutions participating in the \textit{Gaia} Multilateral Agreement.

This research has made use of the NASA/IPAC Infrared Science Archive, which is funded by the National Aeronautics and Space Administration and operated by the California Institute of Technology.

This publication makes use of data products from the Wide-field Infrared 
Survey Explorer, which is a joint project of the University of California, 
Los Angeles, and the Jet Propulsion Laboratory/California Institute of 
Technology, funded by the National Aeronautics and Space Administration.

This research has made use of the VizieR catalogue access tool and the 
SIMBAD database, operated at CDS, Strasbourg, France.

This research has made use of \textsc{topcat} \citep{Taylor2005}.

This research made use of Claude AI (Anthropic) as a programming assistant for the generation and debugging of Python scripts used in figure production and data analysis. It was also used for determination of relevant literature, which was followed up by the author for validation and analysis. All scientific interpretation, data analysis decisions, and conclusions are solely those of the author. All 
AI-assisted content was reviewed and verified by the author prior to inclusion.

\section{Data Availability}
The data underlying this article are available in a Zenodo repository
at \url{https://doi.org/10.5281/zenodo.20149781}. The deposit comprises
three machine-readable CSV files: the kinematic membership catalogue
(\texttt{IC2395\_membership\_173.csv}, 173 members), the disc candidate
catalogue (\texttt{disc\_candidates\_IC2395.csv}, 21 candidates across
the Cross-validated and Single-band tiers), and the combined control
field photometry (\texttt{IC2395\_control\_fields\_combined.csv}), along
with a ReadMe file describing all columns. \textit{Gaia} DR3 data are
publicly available via the \textit{Gaia} Archive
(\url{https://gea.esac.esa.int/archive/}), and AllWISE photometry is
publicly available via the NASA/IPAC Infrared Science Archive
(\url{https://irsa.ipac.caltech.edu}).

\bibliographystyle{aasjournal}
\bibliography{References}

\appendix
\setcounter{figure}{0}
\setcounter{table}{0}
\renewcommand{\thefigure}{{\thesection}\arabic{figure}}
\renewcommand{\thetable}{{\thesection}\arabic{table}}

\section{Full Catalogue of Disc Candidates}
\begin{table*}[h!]
\caption{The complete photometric properties of the 21 identified disc
candidates. $\sigma_{W1\text{-}W2} = \sqrt{\sigma_{W1}^{2} + \sigma_{W2}^{2}}$
is the propagated colour uncertainty; $\Sigma = (W1-W2 - 0.10)/\sigma_{W1\text{-}W2}$
is the excess significance above the 0.10\,mag threshold. This dataset is also
available in machine-readable format via the VizieR Service and Zenodo
(\url{https://doi.org/10.5281/zenodo.20149781}).}
\label{tab:full_appendix}
\makebox[\textwidth][c]{%
\resizebox{\textwidth}{!}{%
\begin{tabular}{lccccccccr}
\hline\hline
Gaia Source ID (DR3) & RA (deg) & Dec (deg) & $G$ (mag) & $W1-W2$ & $\sigma_{W1}$ & $\sigma_{W2}$ & $\sigma_{W1\text{-}W2}$ & $\Sigma$ & Tier \\
\hline
5329217071827655808 & 130.05247 & $-$48.04888 & 17.24 & 0.139 & 0.024 & 0.028 & 0.037 & 1.06 & 1 \\
5329266476819288448 & 129.13440 & $-$47.91230 & 17.86 & 0.108 & 0.026 & 0.031 & 0.040 & 0.20 & 1 \\
5329277029570154368 & 129.28324 & $-$47.93042 & 17.12 & 0.214 & 0.049 & 0.041 & 0.064 & 1.78 & 1 \\
5329301253185618432 & 129.47468 & $-$47.56109 & 17.54 & 0.297 & 0.023 & 0.024 & 0.033 & 5.93 & 1 \\
5329380722948000640 & 129.96001 & $-$47.51929 & 17.93 & 0.601 & 0.025 & 0.022 & 0.033 & 15.04 & 1 \\
5329761055204021888 & 130.95680 & $-$46.74218 & 18.04 & 0.175 & 0.029 & 0.029 & 0.041 & 1.83 & 1 \\
5521710874599156480 & 128.19644 & $-$46.86708 & 17.15 & 0.103 & 0.025 & 0.026 & 0.036 & 0.08 & 1 \\
\hline
5329212158370201472 & 130.10030 & $-$48.17285 & 18.32 & 0.136 & 0.029 & 0.034 & 0.045 & 0.81 & 2 \\
5329265686545254144 & 129.15559 & $-$47.92342 & 18.67 & 0.120 & 0.032 & 0.035 & 0.047 & 0.42 & 2 \\
5329281496336497024 & 129.54838 & $-$47.85065 & 17.88 & 0.103 & 0.029 & 0.033 & 0.044 & 0.07 & 2 \\
5329284515682282624 & 129.53436 & $-$47.82151 & 17.55 & 0.169 & 0.032 & 0.034 & 0.047 & 1.48 & 2 \\
5329286371117405952 & 129.55401 & $-$47.69068 & 17.18 & 0.172 & 0.025 & 0.024 & 0.035 & 2.08 & 2 \\
5329291525069152896 & 129.34398 & $-$47.77301 & 18.25 & 0.157 & 0.027 & 0.027 & 0.038 & 1.49 & 2 \\
5329382921980010752 & 129.76812 & $-$47.53339 & 18.15 & 0.229 & 0.028 & 0.031 & 0.042 & 3.09 & 2 \\
5329386289225840384 & 130.10706 & $-$47.47515 & 17.75 & 0.139 & 0.026 & 0.028 & 0.038 & 1.02 & 2 \\
5329571591300972032 & 131.64941 & $-$47.07248 & 18.71 & 0.158 & 0.033 & 0.048 & 0.058 & 1.00 & 2 \\
5329743252547616640 & 131.11336 & $-$46.97390 & 17.60 & 0.150 & 0.030 & 0.031 & 0.043 & 1.16 & 2 \\
5329819844708154240 & 130.69900 & $-$46.53814 & 18.08 & 0.330 & 0.025 & 0.025 & 0.035 & 6.51 & 2 \\
5521451046263191552 & 129.15212 & $-$47.56702 & 17.81 & 0.175 & 0.027 & 0.030 & 0.040 & 1.86 & 2 \\
5521567246586819584 & 129.64772 & $-$47.02120 & 18.36 & 0.121 & 0.028 & 0.030 & 0.041 & 0.51 & 2 \\
5521588274748427648 & 129.49685 & $-$46.66344 & 18.95 & 0.432 & 0.030 & 0.030 & 0.042 & 7.83 & 2 \\
\hline
\end{tabular}%
}%
}
\end{table*}
\setlength{\tabcolsep}{6pt} 
\clearpage

\section{Membership Catalogue Comparison}
\label{app:comparison}
\setcounter{table}{0}

The 173-member kinematic catalogue was cross-matched against three independent literature membership catalogues for \object{IC\,2395} using a 5\arcsec\ matching radius in TOPCAT. Table~\ref{tab:recovery} summarises the recovery fractions as a function of angular separation from the cluster centre for the two catalogues with sufficient IC\,2395 membership lists for a meaningful radial comparison.

\begin{table}[h!]
\centering
\caption{Recovery fraction of this work's 173 members by \citet{HuntReffert2023} and \citet{Perren2023}, as a function of angular separation from the cluster centre.}
\label{tab:recovery}
\begin{tabular}{lcc}
\hline\hline
Separation & H\&R (2023) & Perren (2023) \\
\hline
$< 0.5\degr$   & 27/28 (96\%) & 21/28 (75\%) \\
$0.5$--$1.0\degr$ & 39/53 (74\%) & 4/53 (8\%) \\
$1.0$--$1.5\degr$ & 27/58 (47\%) & 0/58 (0\%) \\
$1.5$--$2.0\degr$ & 6/34 (18\%)  & 0/34 (0\%) \\
\hline
Total          & 99/173 (57\%) & 25/173 (14\%) \\
\hline
\end{tabular}
\end{table}

Recovery declines monotonically with angular separation for both catalogues, reflecting their more restricted search radii rather than a membership disagreement. Of the 74 members unique to this work (not recovered by \citealt{HuntReffert2023}), 73/74 (99\%) lie beyond 0.5\degr\ from the cluster centre. Nineteen of the 21 disc candidates also lie beyond 0.5\degr\ (separations 0.23--1.95\degr), confirming that the disc-bearing population is concentrated precisely in the region that prior surveys could not access by construction of their search radii.

The overlap with \citet{CantatGaudin2020} is separately discussed. Their catalogue lists 291 \object{IC\,2395} members, but only 16 match the 173 members of this work at 5\arcsec. This low overlap does not arise from a search radius restriction but from a methodological difference: their UPMASK probabilistic algorithm applied to Gaia DR2 selects a different inner-cluster population from the hard kinematic cuts on DR3 applied here, and the two input datasets (DR2 versus DR3) are not directly comparable at the source\_id level.

\begin{figure}[h!]
  \centering
  \includegraphics[width=\textwidth]{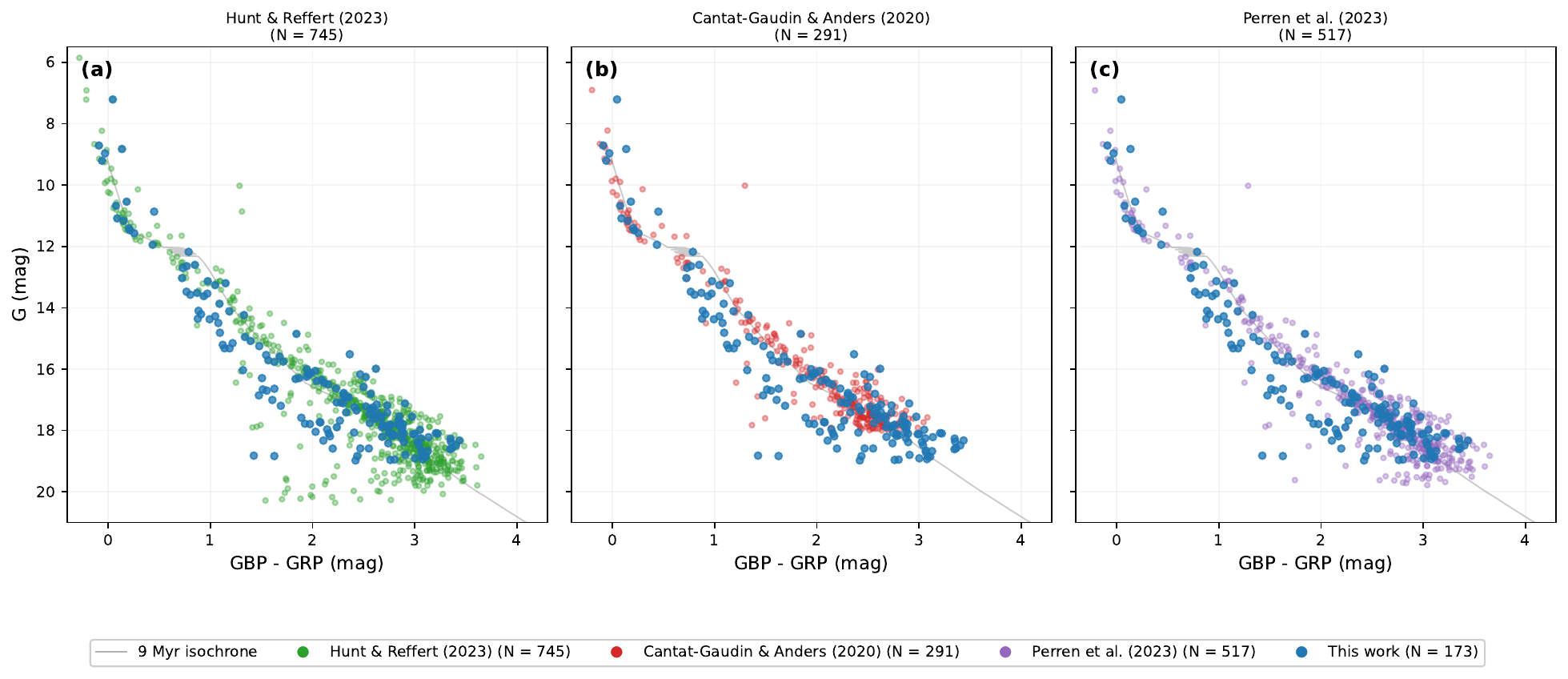}
  \caption{Colour-magnitude diagrams of \object{IC\,2395} kinematic members from three independent literature catalogues compared with the membership catalogue of this work. Panel (a): \citet{HuntReffert2023} ($N = 745$); panel (b): \citet{CantatGaudin2020} ($N = 291$); panel (c): \citet{Perren2023} ($N = 517$). In each panel, the literature catalogue is shown in colour and the 173 members of this work are shown in blue. The faint grey line shows the 9\,Myr PARSEC v2.0 isochrone for reference. The scatter about the stellar sequence is present in all four catalogues, confirming it as a genuine feature of the \object{IC\,2395} stellar population rather than an artefact of any individual membership selection method. The extended faint red population unique to \citet{HuntReffert2023} reflects their more inclusive probabilistic membership threshold relative to the strict kinematic selection employed here.}
  \label{fig:cmd_comparison}
\end{figure}

Residual contamination at large angular separations is constrained by the control field analysis (Section~\ref{sec:control}). The 5.2$\times$ enhancement of disc candidates above the control field background --- matched for Galactic latitude, distance, and nebular environment --- demonstrates that the outer-region members are not dominated by field contaminants. The CMD positions of the 74 unique members are consistent with the 9\,Myr PARSEC isochrone (Figure~\ref{fig:cmd_comparison}), providing additional supporting evidence of physical membership.

\end{document}